\documentclass[aps]{revtex4-2}
\usepackage{amsmath,amssymb}
\usepackage{graphics,graphicx}
\usepackage{dcolumn,bm}
\usepackage{psfrag}
\usepackage{color}

\begin{document}

\title{Virtual walks inspired by a  mean field kinetic exchange model of opinion dynamics}

\author{Surajit Saha$^{1}$, Parongama Sen$^{2}$}%

\address{$^{1}$ Department of Physics, University of Calcutta,
92 Acharya Prafulla Chandra Road, Kolkata 700009, India.\\
$^{2}$ Department of Physics, University of Calcutta,
92 Acharya Prafulla Chandra Road, Kolkata 700009, India.}



\email{psphy@caluniv.ac.in}

\begin{abstract}
We propose two different schemes of realizing a virtual walk 
corresponding to a  
kinetic exchange model of opinion dynamics. The walks are either Markovian or non-Markovian in nature. The opinion dynamics model is characterized by a parameter $p$ which drives an order disorder transition at a critical value $p_c$.  The distribution $S(X,t)$ of the 
displacements $X$ from the origin of the  walkers is computed at different 
times.  
Below $p_c$,  two time scales  associated with a crossover behavior in time are detected,  which diverge in a power law manner at criticality with different exponent values. 
$S(X,t)$  also carries the signature of the phase transition as it changes its form at $p_c$. The walks  show the features of a biased random walk below $p_c$, and  above $p_c$, the walks  are like  unbiased random walks. 
The bias vanishes in a power law manner at $p_c$ and the width of the  resulting Gaussian function shows a discontinuity. 
  Some of the  features of the walks  are argued to be comparable  to the critical quantities  associated with the mean field Ising model, to which class the opinion dynamics model belongs.  
The  results for the Markovian and non-Markovian walks  are almost identical which  is justified by considering the different fluxes.   
We compare the present results with some earlier similar studies. 
 
\end{abstract}




\maketitle
\section{Introduction}

It is a common practice to study  the dynamics of one physical system comprising of some interacting units by mapping
it into another system of interacting particles or pseudo particles in a different space.  
Several such examples can already be found in the literature. 
The correspondence between the zero temperature spin coarsening dynamics in a Ising-Glauber  model in one dimension \cite{privman,derrida1,derrida2,derri2} and
the diffusive motion of the domain walls, which  annihilate as they meet, is well known.
 In the voter model, one can conceive of  a system of coalescing 
walkers which is equivalent to the dynamics of the agents \cite{liggett,krap,howard}
in any dimension.
The dynamics of a opinion formation  model in which the domain sizes play a key role, can also be studied by an equivalent system of annihilating  walkers with a   dynamic bias \cite{sbprps,reshmi}.

On the other hand, one can directly represent the  evolution of  the state of a spin by 
the displacements of a virtual walker in the spin space.   
Such a walk picture has been recently used in \cite{pratik} for a generalized Voter model and earlier,  
several  aspects of this type of a   walk have been  considered 
in different systems  \cite{dornic_g98,drouffe98,newman,balda,luck,drouffe2001}, especially in the context of persistence behavior.
In all these previous studies,  one has a binary spin  variable  and  the spin up (down) state at any time can be thought of as a  displacement of the walker towards right (left) 
in
the one dimensional space  in that particular instant.  
In some of these  works, 
the shape of the probability distribution of the displacements   was shown to have some interesting features; in particular, it carries the signature of the phase transition, if any, occurring in the system.  
 
Such virtual walks  have  been considered previously  
in systems other than spin  or binary opinion dynamics models  also.   Here one may have  more than two  variables in 
the original system that can be 
 mapped to a system of particles with binary states.  For example,   to study the stochastic properties, nucleotide sequences in a DNA
was mapped onto a walk \cite{peng}. For financial data, a random
walk picture was introduced initially  by Louis Bachelier \cite{bachelier}. Later, similar walks
were studied for models of wealth exchange \cite{econo1,econo2}.

 In this paper, we have considered a kinetic exchange model of  opinion dynamics. 
Opinion dynamics models involving kinetic exchange type interactions  have been studied in various ways during the last twenty years or so \cite{deffuant,toscani,lallouache,sb2011,ps2011,sociophysics}. We have considered, in particular, the model proposed in  \cite{soumya2012} and studied the version in which  the opinions can have three discrete values. 
This model  shows a phase transition driven by the disorder in the interaction between the agents. 
Several aspects of this model have been 
studied later \cite{sudip2016,ante2017,oester2019,sbsmps2020}.    

The mean field case, where any agent can interact with any other, has been considered here. 
  We have studied two schemes for the walk, one of them is   Markovian (Scheme I) process and the other non-Markovian (Scheme II). We show that 
the phase transition can be detected from the behavior of the distribution $S(X,t)$, where $X$ is the displacement of the walker at time $t$, for both the walks. The walks show a crossover behavior in time in the ordered phase from which two time scales can be 
defined which diverge close to the critical point with different exponent values. 
We also find that an individual walker is like a biased random walker   below the critical point while above it, the bias vanishes. The behavior of two particular  features of the walks can be argued to be comparable to  those of  the order parameter and specific heat  of the mean field Ising model.

In section II, we have described the  opinion dynamics model under consideration and the related walks in detail.
The results obtained are
presented and analyzed in section III.
In the last section, we  summarize  the work with some conclusive statements.

\section{Description of the opinion dynamics model  and the virtual walks}

We consider the kinetic exchange model (KEM) of opinion dynamics introduced in 
\cite{soumya2012}. However, instead of continuous values, we  take the
case where the  opinions can  take   three discrete values, 0 and  $\pm 1$. 
In this model,  $o_i(t)$, the opinion of the $i$th individual  at time $t$  changes by a  pair-wise interaction in the 
 following manner
\begin{equation}
o_i(t+1)=o_i(t)+\mu_{ij}o_{j}(t). 
\label{dynamics}
\end{equation}
Here $j \neq i$ and 
the choice of pairs is unrestricted, i.e.,  this model is defined on a 
fully connected graph.  $\mu_{ij}$ is the interaction parameter representing the influence of the  $j$th agent  on the $i$th individual. It can take values $\pm 1$;  a negative value is taken with probability  $p$, the only parameter in the model. The opinion value is bounded, 
if it   becomes higher (lower) than $+1$ $(-1)$ then it is made equal to $+1$ $(-1)$. In a system of $N$ agents, the quantity $O=\frac{\mid\sum_i o_i \mid}{N}$ is defined as the order parameter.  
The above mean field model can be exactly solved yielding a 
order-disorder phase transition at $p = p_c = 0.25$, with Ising like criticality\cite{soumya2012}.   



We have associated a walker to each of the individuals of the system in a virtual one dimensional  space. The position of the $i$-th walker at time step $t+1$ in this space  can be written as
\[
X_i(t+1)=X_i(t)+\xi_i(t+1).
\]
At each step the walker can perform one of the three following actions: it can move to the nearest-neighbor site to its right or left  or it can remain at its present location. So, $\xi_i$ is a random number which can take  values $-1$,$0$, or $+1$. 
In this work, we consider the displacements $\xi$ to depend on the opinion 
states. 
We have used two schemes to implement the walk.

Scheme I 
is a Markovian process, i.e. here $\xi_i(t+1)$  depends on the present  opinion states  only: 
\[
\xi_i(t+1)=o_i(t+1). 
\]

Scheme II
is  a non-Markovian walk where the  $\xi_i(t+1)$  depends on the present as well as the previous 
opinion states in the following way:
\begin{eqnarray}
\xi_i(t+1)& =& o_i(t+1)  ~~ {\rm 
if} ~~ o_i(t+1) = o_i(t), \nonumber \\
& = & o_i(t+1)-o_i(t) ~~ {\rm {otherwise}}. \nonumber  
\end{eqnarray}
The values of $\xi$ thus chosen are tabulated in Table \ref{tables1}.
In either case we take $X_i(t=0) = 0$ for all $i$. 
 It is to be emphasized here that the evolution of the opinions directly involves the parameter $p$. The walks on the other hand are solely determined on the basis of the opinions in the last one or two steps and $p$ does not directly enter into the definition of the walk. 

\begin{table}[h]
\begin{center}
\begin{tabular}{ |c|c|c|c|}
\hline
 \multicolumn{2}{|c|} {} &   \multicolumn{2}{c|} {$\xi_i(t+1)$}\\ 			

\cline{1-4}

 $o_i(t)$  & $o_i(t+1)$ &			 Scheme I 	& 	Scheme II	 	 \\

\hline
 1			&	1  		&	 1	& 	1		\\
\hline
 1			&	0  		&	 0	& 	-1		\\
\hline
0 & 0 & 0 & 0\\
\hline
0 & 1 & 1 & 1\\
\hline
0 & -1 & -1 & -1\\
\hline
-1 & 0 & 0 & 1\\
\hline
-1 & -1 & -1 & -1\\
\hline

\end{tabular}
\end{center}
\caption{Table shows the values of $\xi_i(t+1)$ in the two schemes for different values of $o_i$ at times 
$t$ and $t+1$. Note that $|o_i(t) - o_i(t+1)| \leq 1$.}
\label{tables1}
\end{table}

\section{Results}

\begin{figure}
\includegraphics[width=6cm]{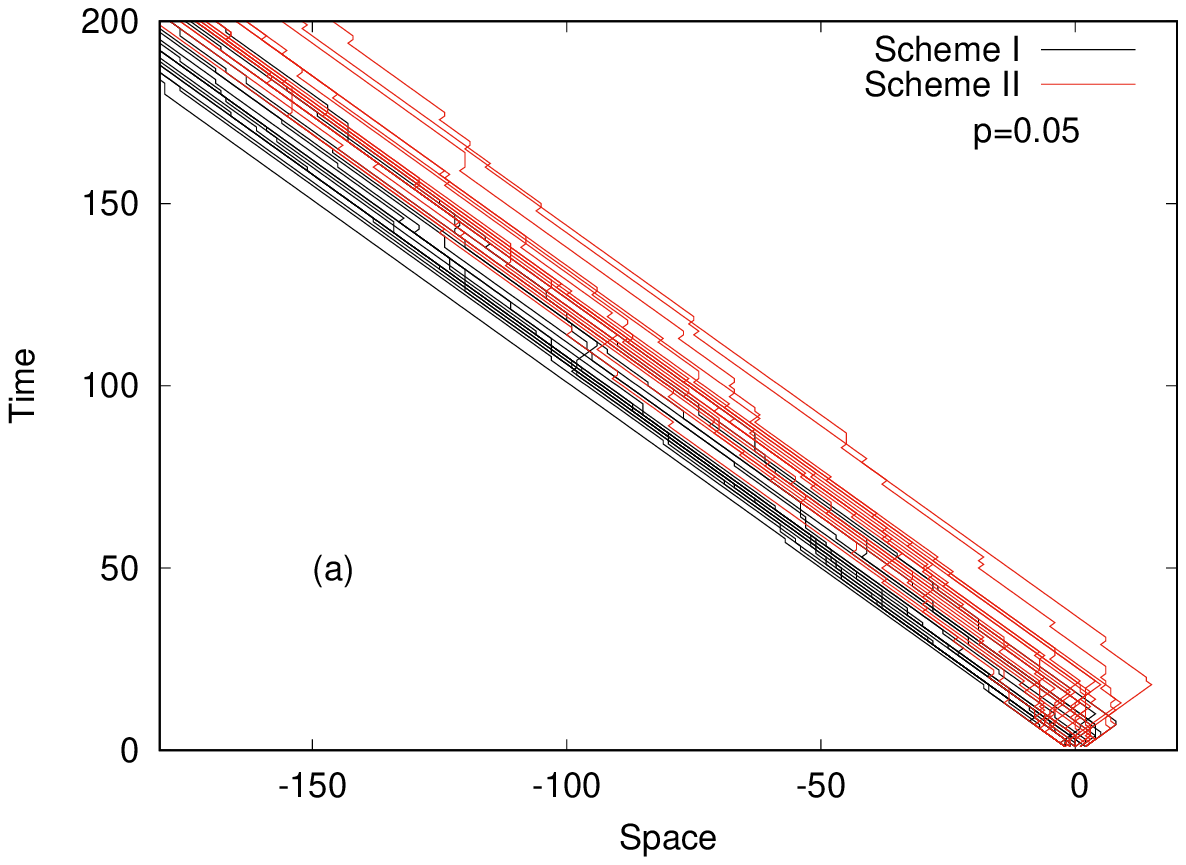}
\includegraphics[width=6cm]{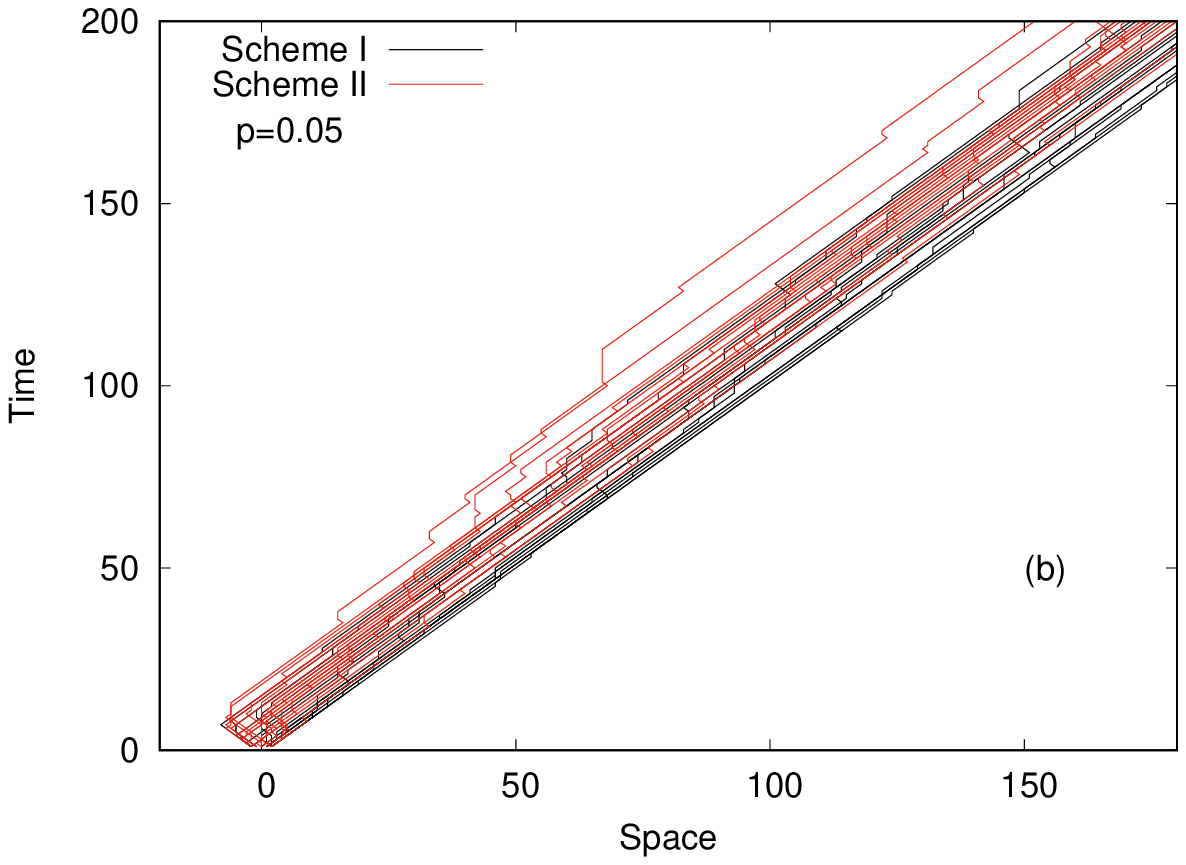}
\includegraphics[width=6.5cm]{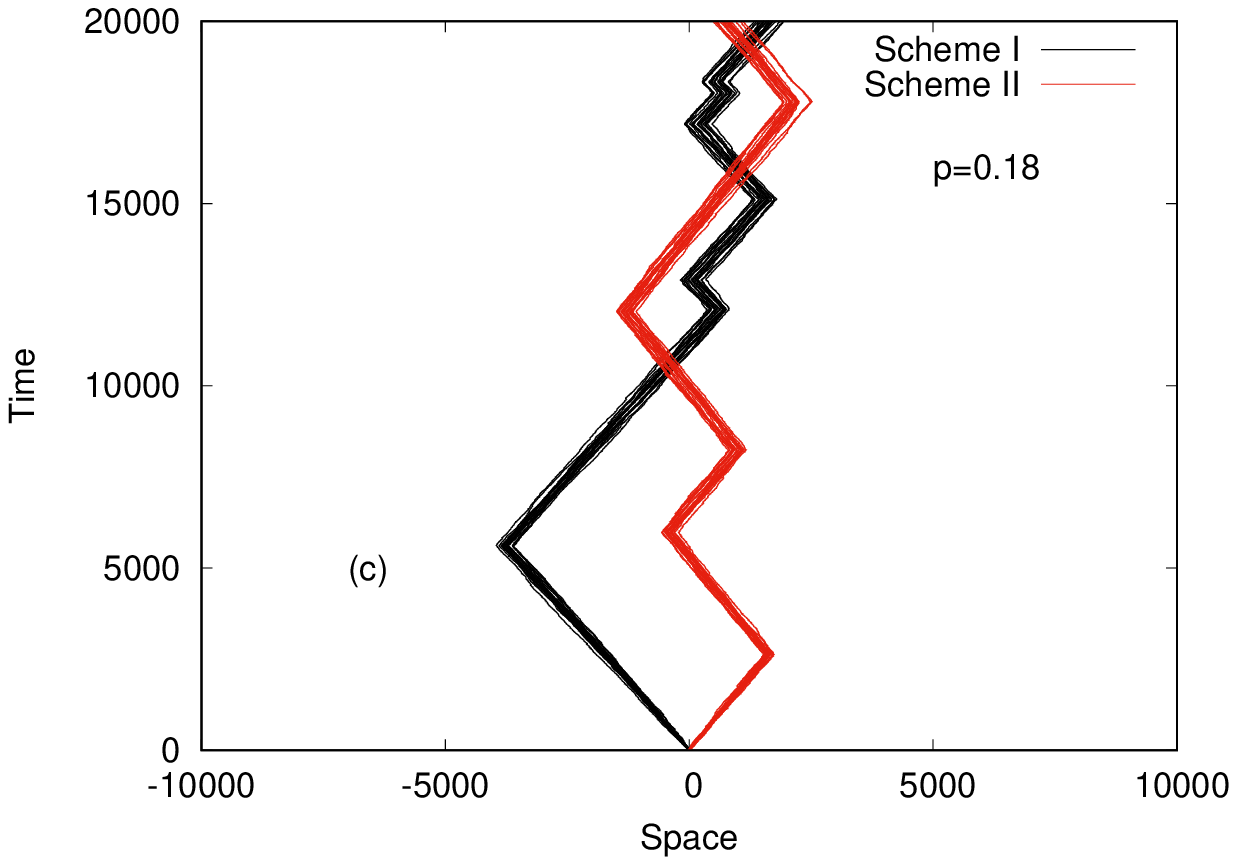}
\includegraphics[width=6.5cm]{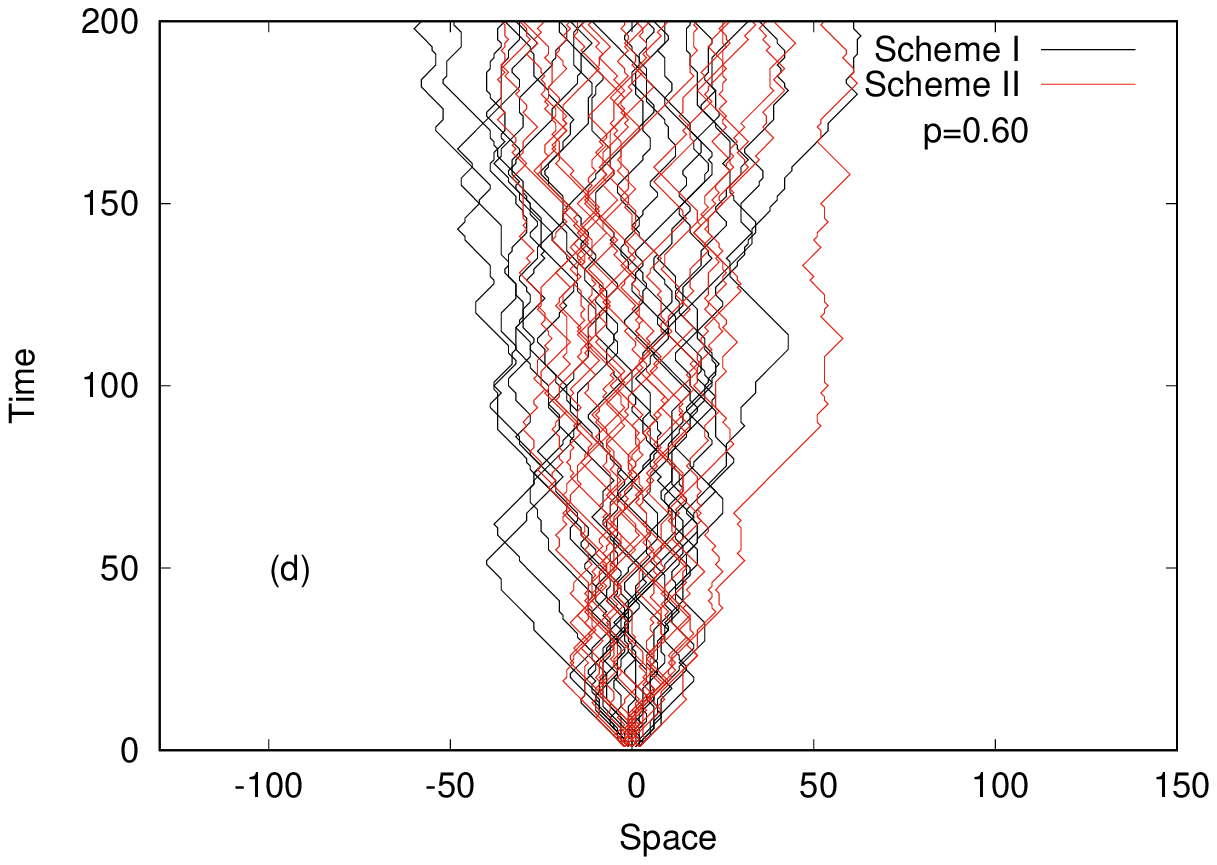}
\caption{(a-c) Snapshots for various values of $p$ shows that for small $p$ and time, the walkers perform more or less ballistic walks. However, changes in direction may be noted at larger times, especially when $p$ is increased. (d) At $p > p_c$, the walks resemble unbiased random walks. The snapshots are shown for both the schemes.}
\label{snap}
\end{figure} 

We have numerically simulated the kinetic exchange model of opinion dynamics on a fully connected graph of size $N$ and started with a completely random
configuration, i.e.,  
at $t=0$, the number of individuals with opinions $0$, $+1$ and $-1$  
is  taken to be $N/3$ for each. 
One Monte Carlo Step (MCS) consists of $N$ updates. 
In each update two different individuals are chosen randomly and    the opinion of the first individual is updated according to eq. \ref{dynamics}. The maximum system size 
simulated is $N=15000$ and the maximum number of configurations over which averaging  has been done is 2000. The  time up to which the model is simulated depends on $N$. 

We first show the world lines  of the walkers as $p$ is varied in  Fig. \ref{snap}. Each of these  snapshots is taken for 20 individuals chosen randomly from a system with   $N=60$.  
It may be noted that for both the schemes,  for small $p < p_c$, 
at very early stages, the walk is centered around zero. At later stages, 
the walkers are more or less directed, either towards the left or right, with 
small diversions in the other direction which lasts for comparatively much shorter 
periods of time as shown in Figs \ref{snap}a and b.   However, for larger $p< p_c$, observations over a longer timescale shows that there are occasional changes in the direction and the walker stays in the same direction over finite intervals of time (see Fig. \ref{snap}c). 
As $p$ exceeds $p_c$, there  is a distinct change in the behavior of the trajectories,  we find that walkers are more or less 
localized with no apparent bias (Fig. \ref{snap}d). 
The picture at $p < p_c$  for larger times, rules out the possibility that the walk is entirely ballistic, as we see occasional directional changes, more so as $p$ increases as indicated in Fig. \ref{snap}. 
 
To gain more quantitative information, the probability distribution $S(X,t)$ for the position $X$ of the walkers at time $t$ is estimated.
The range of $X$ is  $-t \leq X \leq t $.  

\begin{figure}
\vskip -4cm
\includegraphics[width=12cm]{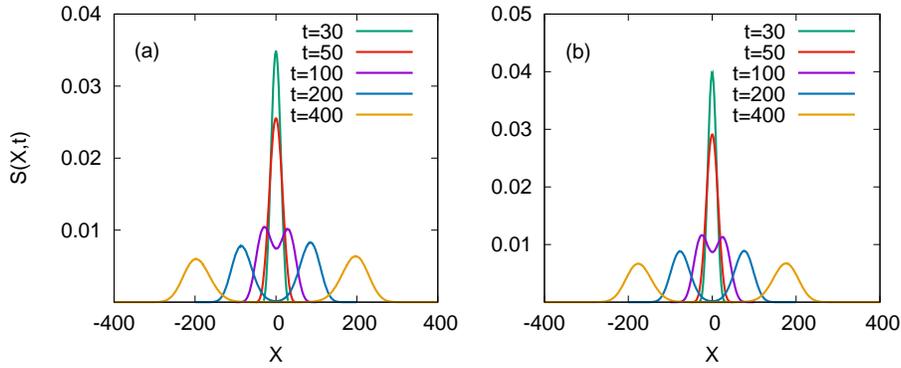}
\caption{Crossover of the probability distribution $S(X,t)$ curves  from a single peaked to a double peaked form for $p=0.2$ in time is shown for $N=15000$ 
in   
 Scheme I (a)  Scheme II (b).}

\label{crossover}
\end{figure} 

\begin{figure}
\includegraphics[width=7cm]{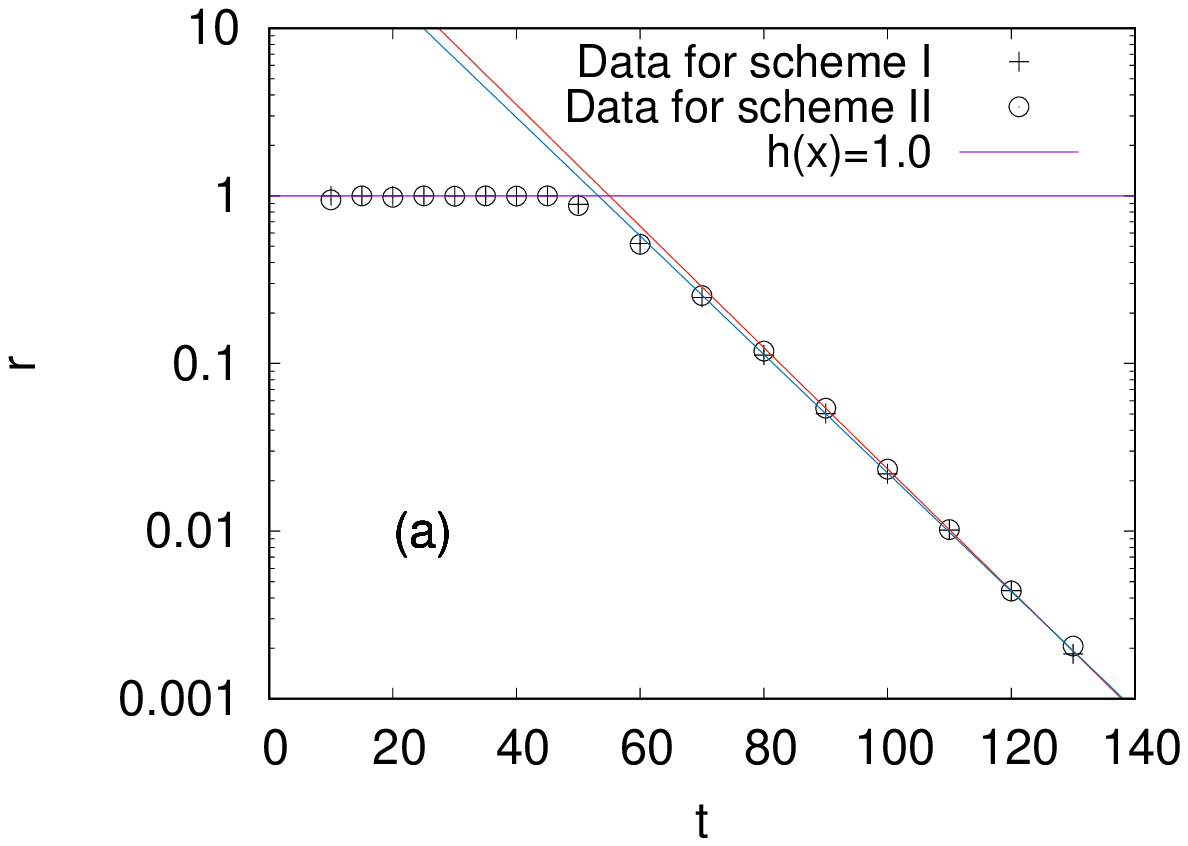}
\includegraphics[width=7cm]{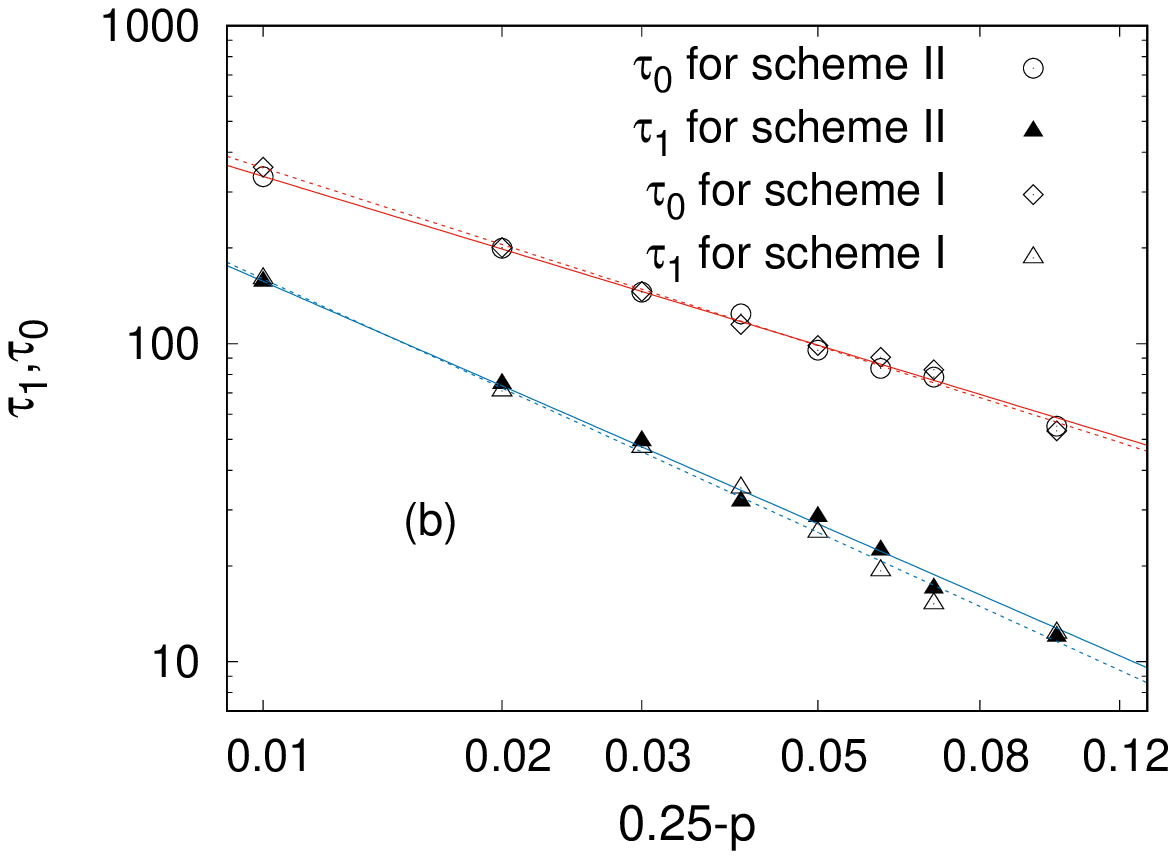}
\caption{(a) Variation of the ratio $r=\frac{S(X=0,t)}{S_{max}(t)}$ with time for $N=15000$ and $p=0.15$  and 
(b) critical behavior of $\tau_0$ and $\tau_1$ shown  for both Scheme I and Scheme II. The 
data can be fit to a power law form as shown and the associated exponents are given in the text and Table \ref{table2}.} 
\label{ratio}
\end{figure}

Before  discussing  the  nature of the distribution, we present the results for a crossover  effect noted in time, for $p < p_c$.  
 During early times, 
$S(X,t)$  shows a single peak at $X=0$ 
but at larger timescales one can observe a crossover from the centrally peaked distribution to a double peaked one. 
The positions of the two peaks are symmetric about the origin. The peaks of $S(X,t)$ occur at larger values of $X$ (Fig. \ref{crossover}) as 
$t$ increases.
To characterize this
crossover behavior,  we estimate the ratio $r = S(X=0,t)/S_{max}(t)$ where  $S_{max}(t)$ is  the maximum value of  $S(X,t)$.  
A study of the variation of $r$ with time 
shows that it remains close to unity till a time $\tau_0$, then vanishes exponentially
for $t > \tau_0$. 
The results for a particular value of $p$ are shown in Fig. \ref{ratio}a.
While $\tau_0$ is the timescale up to which the transient (single peaked) behavior 
of $S(X,t)$ is observed, another timescale $\tau_1$ may be defined from the observation  $r \sim \exp(-t/{\tau_1})$ beyond $\tau_0$.
Fig. \ref{ratio}b shows the variation of the two timescales as a function of $p_c -p$. 
As we approach the critical point $p_c=0.25$, $\tau_0$ and $\tau_1$ both diverge as  
\begin{eqnarray}
\tau_0 & \sim (p_c-p)^{-\mu}, \nonumber \\
 \tau_1 & \sim (p_c-p)^{-\nu}.
\end{eqnarray}
 We thus obtain two new critical exponents $\mu$ and $\nu$;
 $\mu=0.80 \pm 0.01$ for Scheme I and $0.76\pm0.01$ for Scheme II and $\nu$ is equal to $1.14 \pm 0.02 $ for Scheme I and $1.09 \pm 0.02 $ for Scheme II.
We note that  the exponent values differ by less than ten percent for the two schemes.  
The values of the exponents are also shown in Table \ref{table2}.

\begin{figure}
\vskip -4cm
\includegraphics[width=12cm]{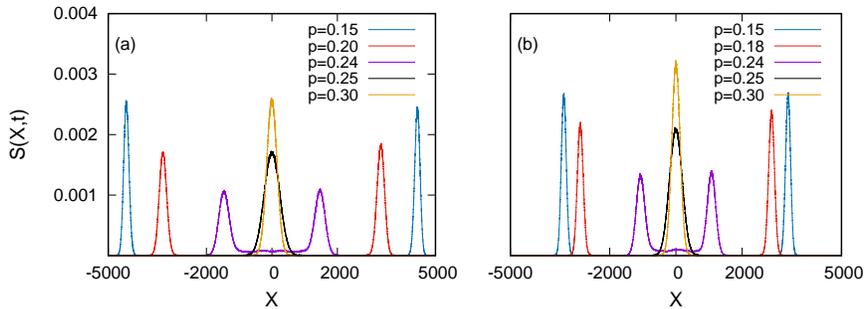}
\caption{$S(X,t)$ curves obtained at $t=5000$ for $N=4000$ for
different values of $p$ in  Scheme I (a) and  Scheme II (b). As the value of $p$ increases, transition
occurs from a double peaked form to single peaked form at $p_c=0.25$ thus signifying the onset of the phase transition.}
\label{prw}
\end{figure}

For $p > p_c$, $S(X,t)$ shows a single peak at the origin and there is no crossover behavior in time. In Fig. \ref{prw}, $S(X,t)$ for different $p$ values at long times shows that the distribution becomes a single peaked one from a double peaked form at $p_c = 0.25$ in both the schemes. Thus we see that the signature of the phase transition is indeed captured in the virtual walks. 

\begin{figure}
\includegraphics[width=6cm]{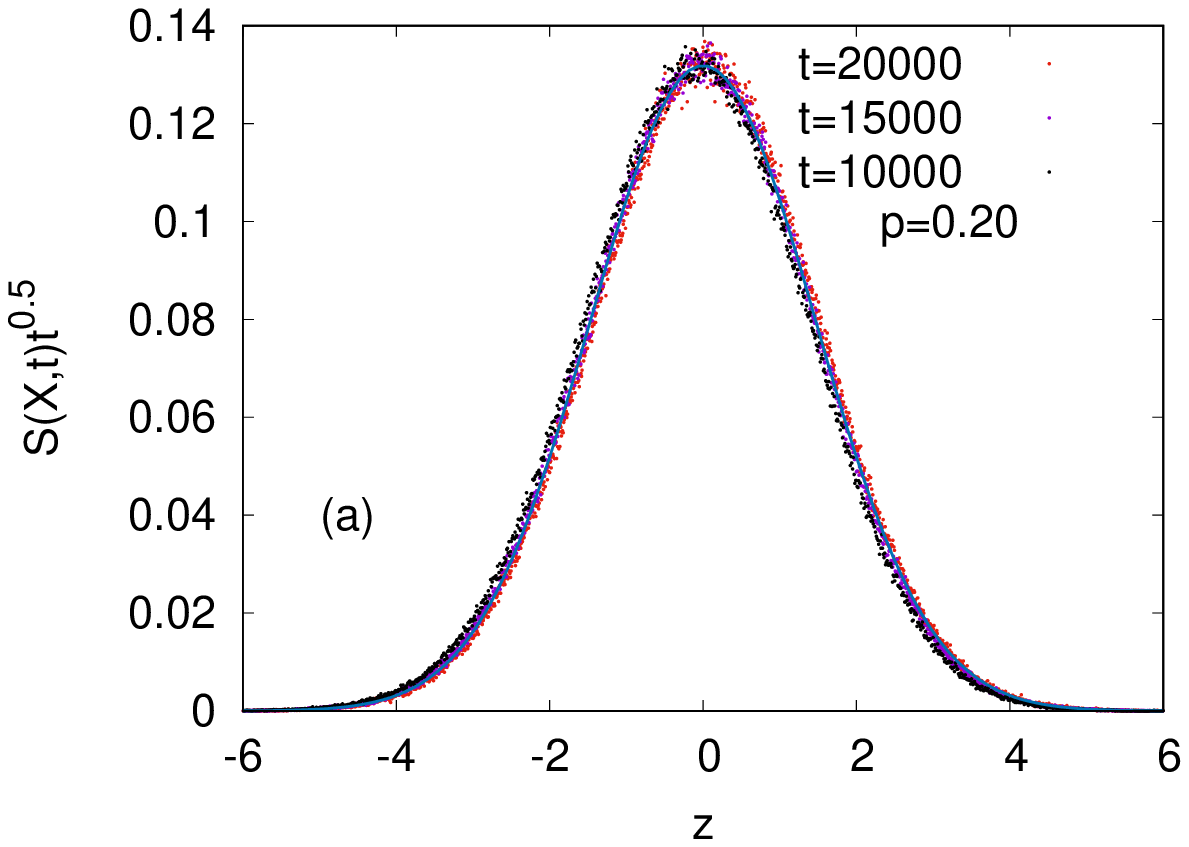}
\includegraphics[width=6cm]{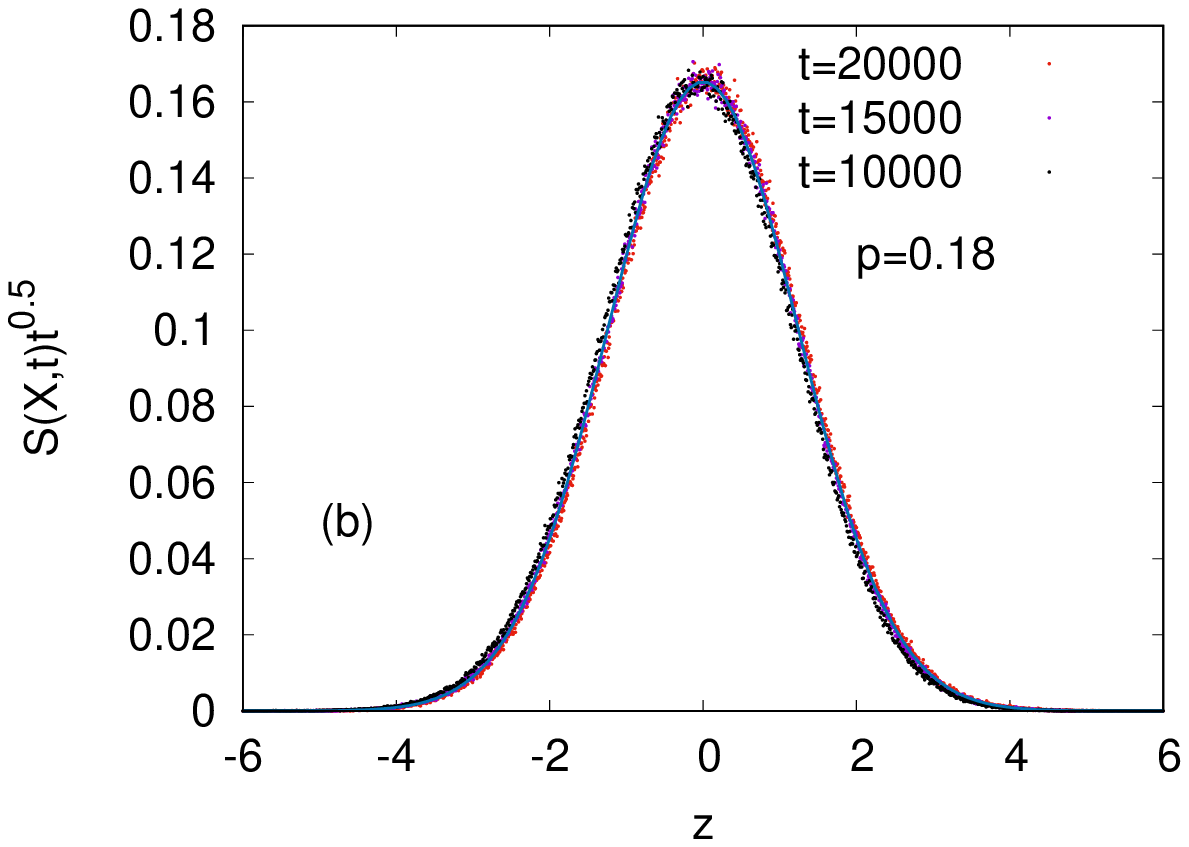}

\includegraphics[width=6cm]{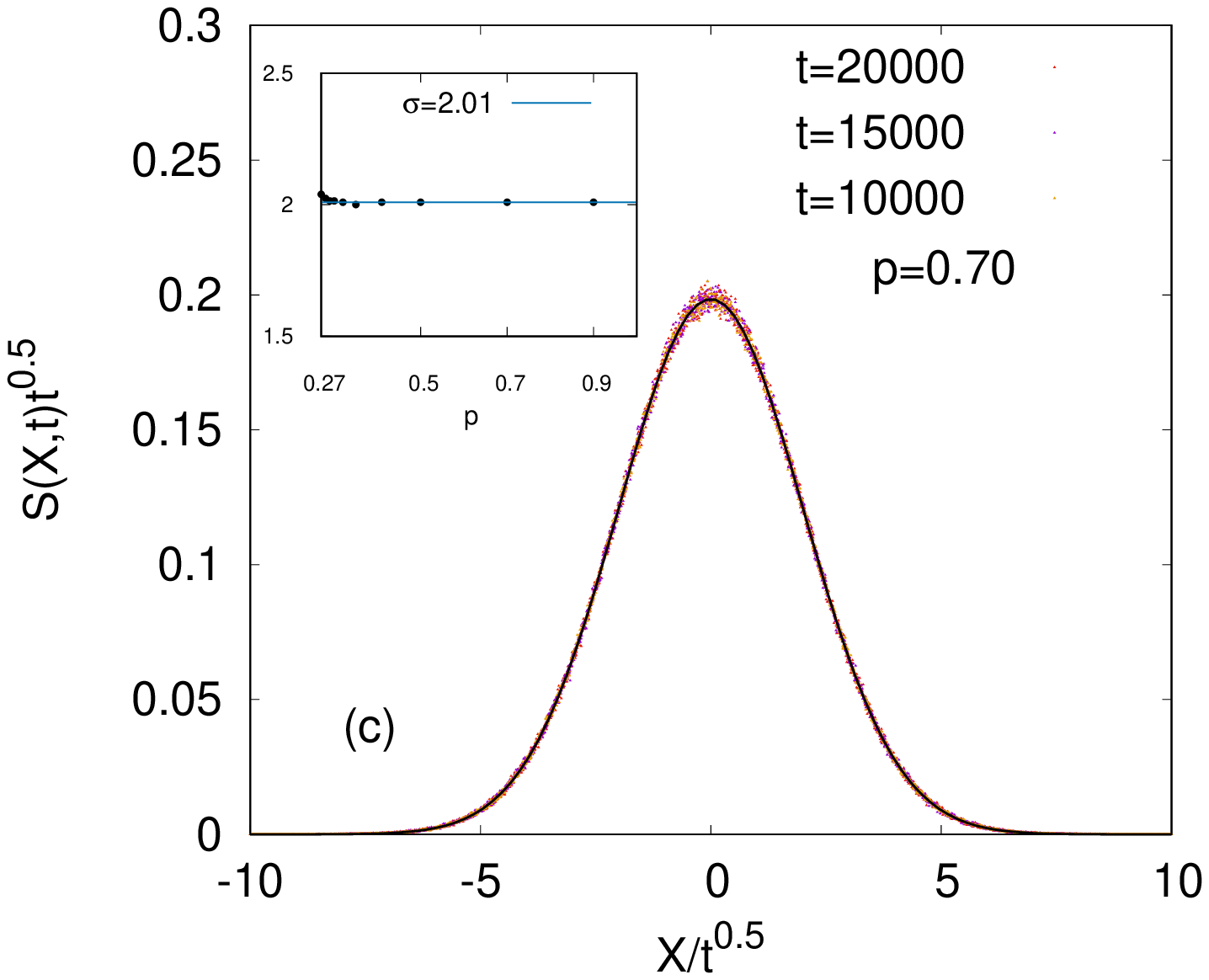}
\includegraphics[width=6cm]{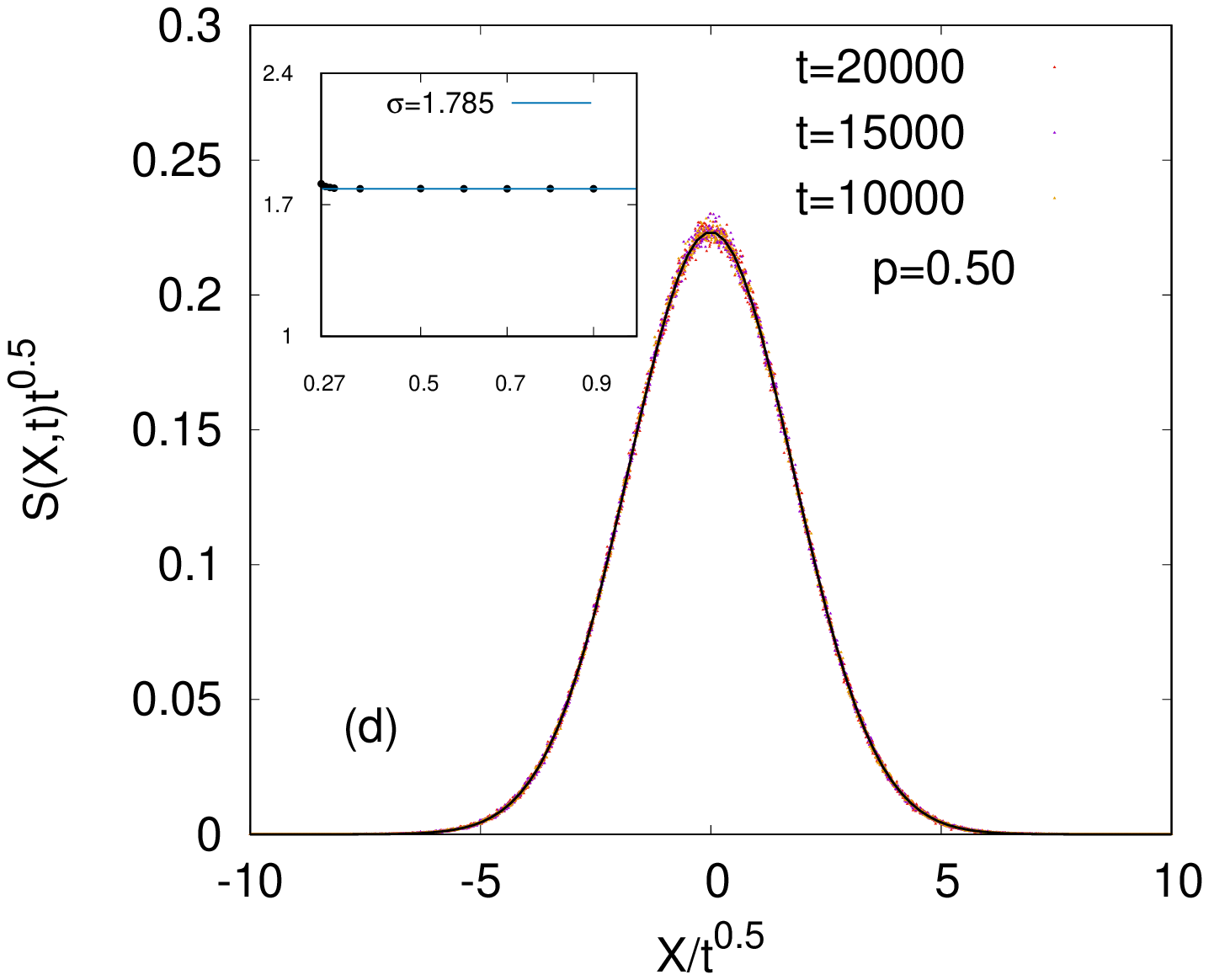}
\caption{Data collapse of $S(X,t)t^{0.5}$ plotted against the scaling  variable $z = \frac{(X - \alpha t)}{\sqrt {t}}$ is shown for $p < p_c$  in  (a) Scheme I ($p = 0.2$) and (b) Scheme II ($p = 0.18$).  (c) and (d) show the data collapse of $S(X,t)t^{0.5}$ 
plotted against $X/t^{0.5}$ 
for $p>p_c$.  
The  insets in (c) and (d) show the  variation of the width  of the Gaussian curves as a function of $p$.  In all the figures (a-d), the fitting Gaussian curves have also been plotted. The system size is $N=4000$ for the above data.
}

\label{scaling}
\end{figure}

We next discuss the form of the distribution beyond the crossover  time.  As already mentioned, for $p < p_c$  there are two peaks occurring symmetrically on both sides of 
the origin. It is therefore  sufficient to focus on, say,  $X > 0$.  
It is natural to have as a first check whether a distribution 
for a walk is random (either biased or unbiased). We indeed find that a nice collapse can be obtained by plotting
$S(X,t)t^{\kappa}$ against $z=\frac{ (X-\alpha t)}{t^{\kappa}}$ where $\kappa= 0.5$ independent of $p$ and $\alpha > 0$ (for $X > 0$) is a function of $p$.  The resulting collapsed curve $f(z)$ is seen to fit to a Gaussian form, i.e., $f(z) \sim \exp(-\frac{z^2}{2\sigma^2})$   where  the width  $\sigma$ depends on $p$. 
For $p > p_c$, we 
find a  similar collapse with $\alpha = 0$. In this regime,  the Gaussian  curve has a width which is nearly a constant except for very close to $p_c$ (which may be a finite size effect). All these results are presented in Fig. \ref{scaling}.

\begin{figure}
\includegraphics[width=7cm]{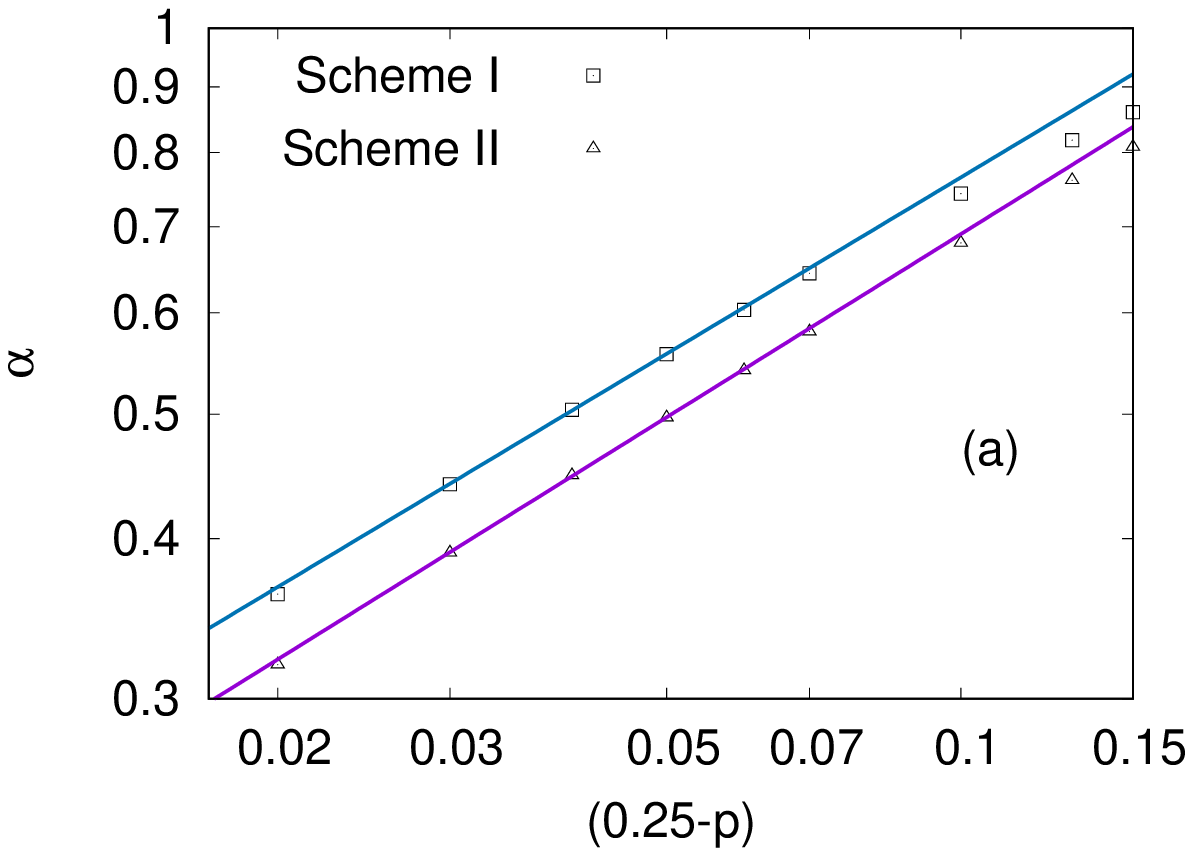}
\includegraphics[width=7cm]{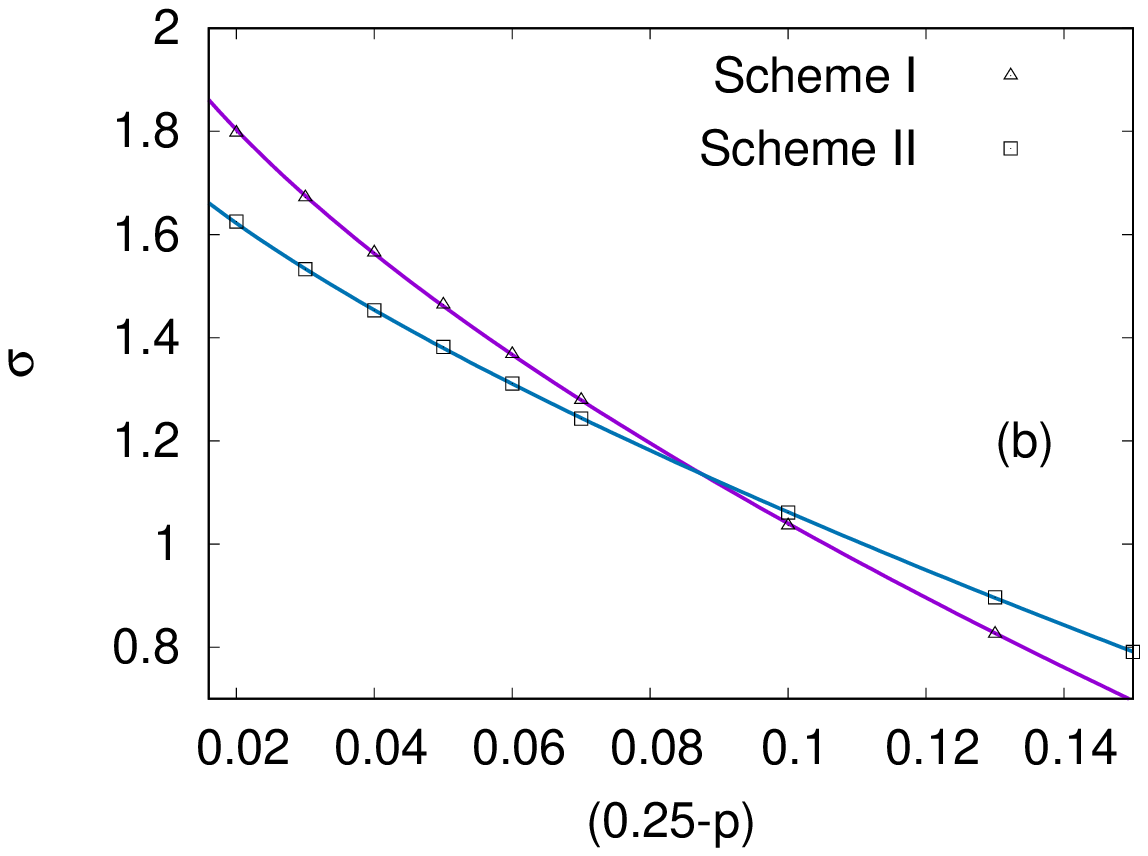}
\caption{Variation of $\alpha$ and $\sigma$ against $p_c-p$ for the two schemes.  The data are fit to the forms given by equations \ref{alpha} and \ref{sigma} as $p \to p_c$. The values of the fitting parameters are shown in Table \ref{table2}. }
\label{parameters1}
\end{figure}

The above results show that the walks are like biased random walks for $p < p_c$ and like an unbiased walk above criticality. 
However, we note that we have here a biased walk where the bias can be  positive and negative, i.e., $\alpha $ will have both positive and negative values.   
It is expected that as $p \to p_c$, $\alpha$ should vanish. This is indeed true and we note by analyzing the data that
\begin{equation}
|\alpha| \sim (p_c - p)^\beta. 
\label{alpha}
\end{equation}
As a function of $p$, $\sigma$ can be  fitted   in the following way:
\begin{eqnarray}
\sigma(p)  &  = & a  -  b(p_c - p)^\gamma ; ~~~ p < p_c \nonumber \\
& = & c~;   ~~~p > p_c 
\label{sigma}
\end{eqnarray}
with $a \neq c$ indicating there is a discontinuity at $p_c$. We will come back to this point later in the last section.
The data for $\alpha$ and $\sigma$ are plotted in Fig. \ref{parameters1} for which the above forms  have been obtained. 

The qualitative behavior discussed above is again true for both the schemes but the values of the parameters differ for the two schemes. We observe that the walk for Scheme II takes place in a more zig-zag manner such that the net displacement is  larger 
for Scheme I. The precise values of the parameters occurring in eqs \ref{alpha} and \ref{sigma} are given in Table \ref{table2}. 

\begin{table}[h]
\begin{center}
\begin{tabular}{|c|c|c|c|c|c|c|c|}
\hline
& $\mu$ & $\nu$  &   $\beta$ & $\gamma$  & $a$& $b$& $c$\\
\hline
Scheme I & $0.80(1)$ & $1.14(2)$ & $0.456(5)$ & $0.61(1)$ & $2.26 (2)$ &$4.99 (7) $ &$\sim 2.01$ \\
\hline
Scheme II & $0.76(1)$  & $1.09(2)$ & $0.474(5)$ & $0.69 (1)$&$1.89(1)  $ &$4.11(4)$ & $\sim 1.78$\\
\hline
\end{tabular}
\end{center}
\caption{Table shows the values of the exponents and the parameters associated with the behavior  of the distribution in the two schemes. }
\label{table2}
\end{table}

\begin{figure}
\includegraphics[width=8cm]{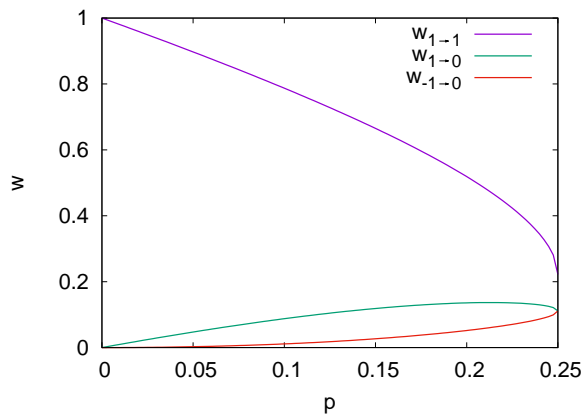}
\caption{The flux  of transition from one state to another 
is shown for three different cases according to equations \ref{rates} and \ref{1-1} when the positive opinions dominate 
in the ordered phase. It is revealed that the flux from a non-zero opinion to a zero
opinion is much less compared to the case where the opinion remains as 1. }
\label{flux}
\end{figure}

  We try to justify next why the two schemes give almost identical results for the exponents $\mu$,  $\nu$ and $\beta$. ($\gamma$ is different for the two walks but it is actually not a critical 
exponent). The fact is, as Table \ref{tables1} shows, the difference 
in the two schemes occurs  only when the opinion state changes from a non-zero value to   zero.  
Note that  below $p_c$, as the order parameter is non-zero, we expect an excess of either opinion value 1 or -1. 
For a configuration where, say, opinion $o=1$ dominates, the expression for the fractions $f_{+}, f_{-}$  of opinion values  $o= \pm1$ can be obtained from \cite{soumya2012} below $p_c$ as
\begin{eqnarray}
f_{\pm} & =  & \frac{(1-3p+2p^2)\pm \sqrt{1-6p+9p^2-4p^3}}{2(1-p)^2},  
\end{eqnarray}
where $f_{+} > f_{-}$. 
Hence the  flux of 
 states with  nonzero opinion to zero opinion   
will be equal  to  
$w_{1 \to 0}$   and $w_{-1 \to 0}$ where 
\begin{eqnarray}
w_{1 \to 0}  & = & f_{+}[ (1-p) f_{-} + p f_{+}], \nonumber \\  
w_{-1 \to 0}  & = & f_{-}[ (1-p) f_{+} + p f_{-}].   
\label{rates}
\end{eqnarray}
The values of these  fluxes are in general less than that of the other fluxes, 
for example,   $w_{1 \to 1}$ is given by
\begin{equation}
w_{1 \to 1} = f_{+} [ \frac{p}{1-p} + (1-p) f_{+} + p f_{-} ].
\label{1-1}
\end{equation}
The first term in eq.\ref{1-1} arises due to the interaction with an agent with opinion value zero, which  occurs with probability $\frac{p}{1-p}$.
In Fig.  \ref{flux},  
the above three fluxes are shown for comparison  that indicates that transition to states with zero 
values are rarer such that the dynamics in  Scheme I and Scheme II  may become nearly identical. 

\section{Summary and discussions}

In this work, we have considered  one dimensional walks in a virtual space corresponding to a kinetic exchange model of opinion dynamics. Two different schemes have been conceived, one Markovian and the other non-Markovian.  The expected feature that the walks contain the signature of the phase transition is obtained. This was also found in \cite{pratik}. 

A crossover behavior in time is also observed and the results show the existence of two timescales  which diverge 
close to the known critical point with  two new exponents. It has been checked that this crossover behavior does not vanish in the thermodynamic limit.  
Such a crossover was also noted in  the voter model \cite{pratik},  although the distribution showed a different form.

The form of the distribution  function shows that the individual walkers perform biased random walk motion where the bias is either to the left or right in the ordered phase. Such a distribution is not noted for similar walks corresponding to binary spin models like the Ising model or generalized voter models 
in two dimensions. 
For these models,  the displacement of the walker was taken to be  simply equal to the spin state. Although Scheme I is a similar walk,  in the present case, since one has a spin zero state as well, the walks are not equivalent which gives rise to the difference in the nature of $S(X,t)$. 
Above the critical point of course, one has a Gaussian walk when the displacements become uncorrelated, independent of the particular model.  

We would also like to add that here we get a biased walk where the bias can be  positive and negative with equal probability on an average. Hence, if one considers the entire distribution,  one will get  $\langle X \rangle =0$ for all $p$. This will give  $\langle X^2\rangle - \langle X \rangle^2  
\propto  t^2$ for $p < p_c$ and  $\propto t$ for $p > p_c$. Note that in ordinary biased random walk, $\langle X^2\rangle - \langle X \rangle^2 \propto t$
in contrast. Using either the region $X> 0$ 
or $X< 0$ we obtained $\langle X^2\rangle - \langle X \rangle^2\propto 
t$ as in the biased random walk.  We argue it is better to regard the  distribution to the left and right of the origin separately, consistent with the  snapshots which 
clearly  reveal  the biased nature of the walks below $p_c$. 
 We also find that the bias $\alpha$ goes to zero above $p_c$ and is therefore analogous to an order parameter. Indeed, the critical exponent associated with 
$\alpha$ is very close to 0.5, the known value for  $O$, the order parameter in the opinion dynamics model (or Ising model) in the mean field case. 

 We also observed a discontinuity in the value of $\sigma$ at $p= p_c$.  In fact the behavior of 
$\sigma$ is quite similar  to that  of the specific heat of the   mean field Ising model with 
the critical exponent equal to zero and a jump discontinuity \cite{stanley}. Note that both  $\sigma$ and specific heat are measures of
fluctuation.  So we claim that from these walks, not only the transition point can be detected, one
can also obtain an estimate of the static critical points. 

As far as the present two schemes are concerned, one gets nearly  identical exponents for the time scales which diverge and the bias that vanishes close to the critical point. A brief analysis helps us to understand why this is so. However, the motion of the walkers is more zig-zag in Scheme II, that leads to a reduced bias and the width of the distribution shows different results quantitatively in the two schemes as a result. In principle, the walks can be conceived in many other ways, so although three exponents show comparable values, one cannot immediately conclude that  a universal behavior exists for all virtual walks. Still, it is interesting that the  Markovian and the  non-Markovian walks show similar behavior.

 \subsection*{Authors' Contributions}
 The work has been  formulated and the manuscript prepared by both authors. Simulations  were  carried out  by SS.

 \subsection*{Funding}
 Research by SS  is  supported by the Council of Scientific and Industrial Research, Government of India through CSIR NET fellowship (CSIR JRF Sanction No. 09/028(1134)/2019-EMR-I). PS has received support from SERB (Government of India) scheme no MAT/2020/000356.




\begin{thebibliography}{99}

\bibitem{privman} Privman V. 1997 \textit{Nonequilibrium Statistical Mechanics in One Dimension} (Cambridge University Press, Cambridge).


\bibitem{derrida1}
Derrida B, Bray AJ and Godr\`eche C. 1994 \textit{Non-trivial exponents in the zero temperature dynamics of the 1D Ising and Potts models} J. Phys. A \textbf{27}, L357.

\bibitem{derrida2}
Derrida B. 1995 \textit{Exponents appearing in the zero-temperature dynamics of the 1D Potts model}. J. Phys. A Math. Theor. \textbf{28}, 1481.

\bibitem{derri2} Derrida B, Hakim V,  Pasquier V. 1995 \textit{Exact First-Passage Exponents of 1D Domain Growth: Relation to a Reaction-Diffusion Model}. Phys. Rev. Lett. \textbf{75}, 751.

\bibitem{liggett}
Liggett TM. 1985 \textit{Interacting Particle Systems} (Springer, New York).

\bibitem{krap} Krapivsky PL, Redner S, Ben-Naim E. 2010 \textit{A Kinetic View of Statistical Physics}. (Cambridge University Press, Cambridge).

\bibitem{howard}
Howard M and Godr\`eche C. 1998 \textit{Persistence in the Voter model: continuum reaction-diffusion approach}. J. Phys. A \textbf{31}, L209.
%
\bibitem{sbprps} Biswas S, Ray P, Sen P. 2011 \textit {Opinion dynamics model with domain size dependent dynamics: novel features and new universality class}. 
Journal of Physics : Conference Series, {\textbf 297}, 012003.  

\bibitem{reshmi} Roy R, Ray P, Sen P. 2020 \textit{ Tagged particle dynamics in one dimensional $A+ A \to kA$ models with the particles biased to diffuse towards their nearest neighbour}. J. Phys. A: Math. Theor. \textbf {53},  155002.

\bibitem{pratik} Mullick P and Sen P. 2018 \textit{Virtual walks in spin space: A study in a family of two-parameter models}.
Phys. Rev. E \textbf{97}, 052122.






\bibitem{dornic_g98}
Dornic I and Godr\`eche C. 1998 \textit{ Mathematical and General
Large deviations and nontrivial exponents in coarsening systems }. J. Phys. A \textbf{31} 5413.

\bibitem{drouffe98}
Drouffe JM and Godr\`eche C. 1998 \textit{
Journal of Physics A: Mathematical and General
Stationary definition of persistence for finite-temperature phase ordering }. J. Phys. A \textbf{31}, 9801.

\bibitem{newman}
Newman TJ and Toroczkai Z. 1998 \textit{Diffusive persistence and the "sign-time" distribution}. Phys. Rev. E \textbf{58}, R2685.

\bibitem{balda}
Baldassarri A, Bouchaud JP,  Dornic I and Godr\`eche C. 1999 \textit{Statistics of persistent events: An exactly soluble model
}. Phys. Rev. E \textbf{59}, R20.

\bibitem{luck}
Godr\`eche C  and Luck JM. 2001 \textit{Statistics of the Occupation Time of Renewal Processes}. J. Stat. Phys. \textbf{104}, 489.

\bibitem{drouffe2001}
Drouffe JM and Godr\`eche C. 2001 \textit{Temporal correlations and persistence in the kinetic Ising model: the role of temperature} Eur. Phys. J. B \textbf{20}, 281.


\bibitem{peng}
Peng CK. 1992 \textit{Long-range correlations in nucleotide sequences}. Nature \textbf{356}, 168.

\bibitem{bachelier}
Bachelier L. 1900 \textit{Theorie de la speculation}, Annales Scientifiques de I'Ecole Normale Superiure, \textbf{3} (17), pp. 21-86.

\bibitem{econo1}
Chatterjee A and Sen P. 2010 \textit{Agent dynamics in kinetic models of wealth exchange}. Phys. Rev. E \textbf{82}, 056117.

\bibitem{econo2}
Goswami S, Chatterjee A and Sen P. 2011 \textit{Antipersistent dynamics in kinetic models of wealth exchange}.  Phys. Rev. E \textbf{84}, 051118.


\bibitem{deffuant} Deffuant G, Neau D, Amblard F, Weisbuch G. 2000 \textit{Mixing beliefs among
interacting agents}. Advances in Complex Systems, \textbf{3}, 87.
\bibitem{toscani} Toscani G. 2006 \textit{Kinetic models of opinion formation}. Communications in Math-
ematical Sciences,\textbf{4}, 481.


 \bibitem{lallouache}  Lallouache M, Chakrabarti A.S, Chakraborti A, and Chakrabarti B.K. 2010\textit{
Opinion formation in kinetic exchange models: Spontaneous symmetry-breaking
transition}. Phys. Rev. E, \textbf{82},056112.

\bibitem{ps2011} Sen P. 2011 \textit{Phase transitions in a two-parameter model of opinion dynamics with random kinetic exchanges}. Phys. Rev. E \textbf{83}, 016108.

\bibitem{sb2011} Biswas S. 2011 \textit{Mean-field solutions of kinetic-exchange opinion models}. Phys. Rev. E \textbf{84}, 056106.
\bibitem{sociophysics} Sen P, Chakrabarti BK. 2014  \textit{B.K.Sociophysics: An Introduction}(Oxford University Press, Oxford) and the references therein.

\bibitem{soumya2012} Biswas S, Chatterjee A, Sen P. 2012 \textit{Disorder induced phase transition in kinetic models of opinion dynamics}. Physica A \textbf{391}, 3257.


\bibitem{sudip2016} Mukherjee S, Chatterjee A. 2016 \textit{Disorder-induced phase transition in an opinion dynamics model: Results in two and three dimensions}. Phys. Rev. E \textbf{94}, 062317.

\bibitem{ante2017} Anteneodo C, Crokidakis N. 2017 \textit{Symmetry breaking by heating in a continuous opinion model}. Phys. Rev. E \textbf{95}, 042308.

\bibitem{oester2019} Oestereich AL, Pires MA, Crokidakis N. 2019 \textit{Three-state opinion dynamics in modular networks}. Phys. Rev. E \textbf{100}, 032312.


\bibitem{sbsmps2020} Mukherjee S, Biswas S, Sen P. 2020 \textit{Long route to consensus: Two-stage coarsening in a binary choice voting model}. Phys. Rev. E \textbf{102}, 012316. 


 


\bibitem{stanley} Stanley HE. 1971 \textit{Introduction to Phase Transitions and Critical Phenomena}(Oxford University Press, London).


\end{thebibliography}
\end{document}